\documentclass[prd,twocolumn,superscriptaddress,floatifx,amsmath,amssymb,amsfonts,nofootinbib]{revtex4-1}

\usepackage{float} 

\usepackage[normalem]{ulem}
\usepackage[english]{babel}
\usepackage{graphicx}
\usepackage{dcolumn}
\usepackage{bm}
\usepackage{blindtext}
\usepackage{verbatim}
\usepackage{relsize}
\usepackage{mathrsfs}
\usepackage{musicography}
\usepackage{amsmath}
\usepackage{cancel}
\usepackage{physics}
\usepackage{epstopdf}
\usepackage{mathtools}
\usepackage{tensor}
\usepackage{color}
\usepackage[usenames,dvipsnames]{pstricks}
\usepackage{epsfig}
\usepackage{pst-grad} 
\usepackage{pst-plot} 
\usepackage{hyperref}
\usepackage{verbatim}
\usepackage{slashed}

\newcommand{\mf}{\mathsf}

\allowdisplaybreaks[1] 

\begin{document}

\title{General features of the thermalization of particle detectors and the Unruh effect}

\author{T. Rick Perche}
\email{trickperche@perimeterinstitute.ca}
\affiliation{Perimeter Institute for Theoretical Physics, Waterloo, Ontario, N2L 2Y5, Canada}
\affiliation{Department of Applied Mathematics, University of Waterloo, Waterloo, Ontario, N2L 3G1, Canada}

\begin{abstract}
    We study the thermalization of smeared particle detectors that couple locally to \emph{any} operator in a quantum field theory in curved spacetimes. We show that if the field state satisfies the KMS condition with inverse temperature $\beta$ with respect to the detector's local notion of time evolution, reasonable assumptions ensure that the probe thermalizes to the temperature $1/\beta$ in the limit of long interaction times. Our method also imposes bounds on the size of the system with respect to its proper acceleration and spacetime curvature in order to accurately probe the KMS temperature of the field. We then apply this formalism to a uniformly accelerated detector probing the Minkowski vacuum of any CPT symmetric quantum field theory, and show that the detector thermalizes to the Unruh temperature, independently of the operator it couples to. This exemplifies yet again the robustness of the Unruh effect, even when arbitrary smeared detectors are used to probe general operators in a quantum field theory.
\end{abstract}

\maketitle

\section{Introduction}
    
    Quantum field theory is among the most successful theories of physics. It has given the most accurate predictions for microscopic effects that have ever been measured. However, not all predictions of quantum field theory have been experimentally confirmed. Among the most general predictions of the theory is the Unruh effect, which claims that a uniformly accelerated observer with proper acceleration $a$ experiences a thermal bath of temperature $T_U$, proportional to its acceleration,
    \begin{equation}
        T_{U} = \frac{\hbar a}{2\pi c k_B}.
    \end{equation}
    \textcolor{black}{Named after Unruh due to his approach in~\cite{Unruh1976}}, and later explored by many~\cite{fullingUnruhEffect,Davies1974,Unruh-Wald,takagi,waitUnruh,edunruh,unruhEffectNoThermal,unruhSlow}, the Unruh effect is a prediction of quantum field theory that is able to connect three of the most fundamental areas of physics: quantum mechanics, relativity and thermodynamics, as can be seen by the appearance of the constants $\hbar$, $c$ and $k_B$ in the equation above.
    
    There are many different methods that can be used to approach the Unruh effect~\cite{Davies1974,fullingUnruhEffect,Bisognano1975,Bisognano1976,Unruh1976,Unruh-Wald,takagi,matsasUnruh,waitUnruh,edunruh,unruhEffectNoThermal,unruhSlow}, all of which agree on the value of the Unruh temperature. One possible way to describe this effect relies on the Bogoliubov coefficients that relate the Minkowski modes and the so called Rindler modes~\cite{Davies1974,fullingUnruhEffect,takagi,matsasUnruh}. This decomposition shows that the Minkowski vacuum is a two-mode squeezed state with respect to the modes on the two Rindler wedges. This elementary description shows that when restricting the state to one of the wedges, it yields a thermal state with inverse temperature $\beta_U = 1/T_U$. Another way of obtaining the conclusion regarding the Unruh effect, this time involving formal aspects of algebraic quantum field theory, involves showing that the Minkowski vacuum is a KMS state with respect to time evolution along the Rindler wedge with inverse temperature $\beta_U$~\cite{Bisognano1975,Bisognano1976,sewell,UnruhPhilosophers}.
    
    Although the approaches mentioned above are more than enough to argue in favour of the Unruh effect, there is yet another way of probing it, and it is perhaps the most physical one of them. This method consists on the use of particle detectors~\cite{Unruh1976,Unruh-Wald,takagi,waitUnruh,edunruh,unruhEffectNoThermal,unruhSlow,garay}. A particle detector is a non-relativistic quantum system that couples locally to an operator in the algebra of observables of a quantum field theory. The way to use a particle detector to probe the Unruh effect is rather simple: one prescribes a uniformly accelerated trajectory for the detector and considers an interaction when the field state is the Minkowski vacuum. The result after a long time interaction is that the detector ends up in a thermal state with temperature $T_U$. 
    
    Although particle detector models were introduced by Unruh for the first time in order to better understand black hole thermodynamics and the related Unruh effect, their use has broaden ever since. In fact, it has been shown that particle detector models can be been used to model multiple physical scenarios, such as the interaction of light with matter~\cite{Pozas2016,eduardoOld,eduardo,Nicho1,richard} and neutrino detection~\cite{neutrinos,antiparticles}. Furthermore, from the perspective of relativistic quantum information, particle detectors allow one to apply quantum information protocols to relativistic quantum field theories, allowing one to harvest entanglement from quantum fields in flat and curved spacetimes~\cite{Valentini1991,Reznik2003,Reznik1,Pozas-Kerstjens:2015,Pozas2016,Nick,topology} and to perform innovative techniques, such as quantum energy teleportation~\cite{masahiro08,masahiro09}.
    
    In this manuscript we are concerned with the thermalization of particle detectors when they move along trajectories such that the field state is a thermal state of inverse temperature $\beta$ with respect to the detector's motion. We consider a smeared particle detector interacting with a general tensor valued field operator and show that under a rigidity assumption for the detector and an adiabatic switching of the interaction, it is possible to measure the temperature of the field state after long interaction times, regardless of the detector's shape and internal degrees of freedom. This result holds for general trajectories (even in curved spacetimes) and is independent of the field operator that the detector couples to. In particular, our results show that any rigid sufficiently localized uniformly accelerated quantum system that couples to a quantum field theory experiences the Unruh thermal bath with temperature $T_U$.
    
    This manuscript is organized as follows: In Section \ref{secKMS} we briefly review the KMS condition in quantum mechanics and quantum field theory. Section \ref{secFNC} is devoted to a brief review of Fermi normal coordinates and the local notion of time evolution associated with a timelike trajectory. In Section \ref{secPD} we define the general smeared particle detector model we will work with and study its basic properties. In Section \ref{secMain}, we show that it is possible to recover the temperature of the field state by means of the excitation deexcitation ratio and exemplify our formalism in the case of a uniformly accelerated particle detector. The conclusions of this work can be found in Section \ref{sec:Conclusions}. 

\section{The KMS Condition and Thermality}\label{secKMS}
    
    Ever since its formulation in the early 20th century, quantum mechanics has proven to be a very successful theory. Few yeas after the first developments of the formalism, Klein, Gordon, Dirac and others~\cite{Klein,Gordon,dirac,QED} were able to reconcile quantum theory with the theory of special relativity that had been recently formulated by Einstein by creating the first relativistic quantum field theories. However, it wasn't until the middle of the century that the most meaningful contributions to conciliate quantum mechanics and thermodynamics were made. In fact, the most general notion of thermal states in quantum mechanics was only introduced by Kubo, Martin and Schwinger in the late 50s~\cite{Kubo,MartinSchwinger}. In this section we review such notion of thermality in non-relativistic quantum mechanics and its extension to relativistic field theories.
    
    \subsection{KMS states as a generalization of Gibbs states}
    
    In this section we review the concept of temperature for states in thermal equilibrium in quantum mechanics. The simplest description of a quantum state in thermal equilibrium is in terms of Gibbs states. Assume a quantum system to be defined in a Hilbert space $\mathcal{H}$ with a Hermitian time independent Hamiltonian $\hat{H}$, whose spectrum is discrete and bounded from below. Then the Gibbs state with inverse temperature $\beta$ is defined as the density operator
    \begin{equation}
        \hat{\rho} = \frac{e^{-\beta \hat{H}}}{\mathcal{Z}},
    \end{equation}
    where $\mathcal{Z}$ is the partition function,
    \begin{equation}\label{Z}
        \mathcal{Z} = \tr(e^{-\beta\hat{H}}).
    \end{equation}
    Notice that the definition of a Gibbs state relies on regularity conditions regarding the Hamiltonian. In fact, in more general setups (such as most quantum field theories), Gibbs states are ill defined. Nevertheless, 
    there is a more general definition of thermality that relies on properties of the expected values of operators in thermal states. This more general definition of thermality that carries to more general scenarios is the so called  Kubo-Martin-Schwinger (KMS) condition, which defines the KMS states~\cite{Kubo,MartinSchwinger}. 
    
    In its essence, the KMS condition provides a relationship between periodicity in imaginary time and temperature. Thus, it is natural that in order to define the KMS condition, we need the notion of time evolution in quantum mechanics. Time evolution can be formulated in terms of a \textcolor{black}{strongly continuous} one-parameter family of unitary operators $U:\mathbb{R}\rightarrow \mathcal{B}(\mathcal{H})$, where $\mathcal{B}(\mathcal{H})$ denotes the set of bounded operators in the Hilbert space $\mathcal{H}$~\cite{KMSreview,Haag}. In the Heisenberg picture, the operators of the quantum theory evolve in time according to $\hat{U}(t)$ as
    \begin{equation}
        \hat{A}(t) = \hat{U}^\dagger(t) \hat{A} \hat{U}(t).
    \end{equation}
    The application of $\hat{U}(t)$ to operators as above defines a one parameter family of automorphisms, \mbox{$\alpha_t:\mathcal{L}(\mathcal{H}) \rightarrow \mathcal{L}(\mathcal{H})$} that maps a linear operator $\hat{A}$ to its time evolution at time $t$, $\hat{A}(t)$.
    
    When the Hamiltonian of the system is time independent, the time evolution operator can be written in terms of the Hamiltonian of the system as
    \begin{equation}
        \hat{U}(t) = e^{-i\hat{H}t}.
    \end{equation}
    Under the assumption that the state of the system $\hat{\rho}$ is a Gibbs state at inverse temperature $\beta$, it is not hard to show that for any bounded operators $\hat{A}$ and $\hat{B}$ the following condition holds,
    \begin{equation}\label{KMSQM}
        \ev{\alpha_t(\hat{A})\hat{B}}_{\hat{\rho}} = \ev{\hat{B}\,\alpha_{t+i\beta}(\hat{A})}_{\hat{\rho}},
    \end{equation}
    whenever the two-point correlator $\langle\alpha_t(\hat{A})\hat{B}\rangle_{\hat{\rho}}$ is an analytical function of $t$ in the complex strip $0<\Im(t)<\beta$. Conversely, if the partition function in Eq. \eqref{Z} is well defined, it is possible to show that a state that satisfies Eq. \eqref{KMSQM} for every pair of bounded operators $\hat{A}$ and $\hat{B}$ is a Gibbs state at inverse temperature $\beta$, provided that the two-point correlator is an analytical function in the complex strip $0<\Im(t)<\beta$. However, even in cases where the partition function is not well defined
    , Eq. \eqref{KMSQM} still makes sense and it can be used to define thermal states. In fact, a KMS state at inverse temperature $\beta$ is defined as a state that satisfies condition \eqref{KMSQM} for \textcolor{black}{every pair of bounded} operators $\hat{A}$ and $\hat{B}$ and such that the correlator is an analytical function for $0<\Im(t)<\beta$. It is then a generalization of the notion of Gibbs states that can be applied to more general scenarios.

    \subsection{The KMS condition in quantum field theory}
    
    In the context of quantum field theory, the KMS condition can be used to provide a notion of thermality with respect to a given notion of time translation. However, an extra difficulty shows up in relativistic theories due to the fact that the operators are defined locally in spacetime, and therefore, in order to define the one-parameter family of automorphisms $\alpha_\tau$ that shows up in Eq. \eqref{KMSQM}, one must first define a notion of time translation. Our approach will be to associate a notion of time evolution with a timelike vector field $\zeta$ defined in an open region $\mathcal{O}$ of the spacetime $\mathcal{M}$ that contains its support. The flux of $\zeta$ by a parameter $\tau$, $\Phi_\tau:\mathcal{O}\rightarrow \mathcal{O}$ is a local diffeomorphism within $\mathcal{O}$ and defines a local notion of time translation. \color{black} We will then see how this notion can induce a one parameter family of time evolution automorphisms $\alpha_\tau$, where $\tau$ should be understood as the flux parameter associated with $\zeta$.
    
    A formal formulation of the KMS condition in QFT is given by the algebraic quantum field theory (AQFT) approach. In this approach, the QFT is defined in terms of an association of open sets in the spacetime $\mathcal{M}$ with *-algebras~\cite{Haag,kasia}. These algebras can then be represented as sets of operators in a Hilbert space according to a GNS representation~\cite{Haag,kasia}. 
    One way of building the *-algebras associated to open sets in spacetime is by formally defining operator valued distributions~\cite{kasia,aqft}. In the case of a real scalar QFT, what is usually known as the quantum field, $\hat{\phi}(\mf x)$, can be seen as an operator valued distribution that formally acts on compactly supported functions $f(\mf x)$ according to
    \begin{equation}\label{smearedPhi}
        \hat{\phi}(f) = \int \dd V f(\mf x) \hat{\phi}(\mf x),
    \end{equation}
    where $\dd V$ is the invariant spacetime volume element. In this sense, in the AQFT approach, although $\hat{\phi}(\mf x)$ is ill defined, the association $f\mapsto \hat{\phi}(f)$ can be used to define the generators of the algebra of operators associated with the quantum field theory.
    
    
    
    We consider a quantum field theory in a spacetime $\mathcal{M}$ such that the operators of the theory can be seen as operator valued distributions, in a similar way to the scalar case. In fact, it can be shown that spinor and vector field theories can be built from a similar construction~\cite{aqft}. Let $\hat{O}(\mf x)$ be any operator valued distribution in this QFT. In particular, $\hat{O}(\mf x)$ might be a tensor valued operator, $\hat{O}_a(\mf x)$, where $a$ stands for any collection of Lorentz indices. Then, $\hat{O}_a(\mf x)$ formally acts on compactly supported test fields $f^a(\mf x)$, conjugate to $\hat{O}(\mf x)$, according to
    \begin{align}\label{Of}
        \hat{O}(f) = \int \dd V f^a(\mf x)\hat{O}_a(\mf x).
    \end{align}
    An example of such tensor valued distribution is, for instance, $\nabla_\mu \hat{\phi}(\mf x)$, where $\hat{\phi}(\mf x)$ is a real scalar field and $\nabla$ is the connection in $\mathcal{M}$. $\nabla_\mu\hat{\phi}(\mf x)$ then formally acts on compactly supported vector fields $f^\mu(\mf x)$.
    
    Under the assumption that the operators of the quantum field theory under consideration can be understood in the sense of Eq. \eqref{Of}, the action of the one parameter family of automorphisms $\alpha_\tau$ can be defined in terms of the time translation induced by $\zeta$. We define the action of $\alpha_{\tau}$ on $\hat{O}(\mf x)$ according to
    \begin{equation}
        \alpha_{\tau}(\hat{O})(f) \coloneqq \hat{O}((\Phi_{\tau})_* f) = \int \dd V ((\Phi_{\tau})_* f)^a(\mf x)\hat{O}_a(\mf x),
    \end{equation}
    where $(\Phi_{\tau})_*$ denotes the pushforward\footnote{An intuitive way of understanding the action of $\alpha_{\tau}$ on $\hat{O}_a$ is by noting that it acts analogously to the pullback on forms, in the sense that if $\omega$ is a one-form and $v$ is a vector field, then the pullback of $\omega$ acts on $v$ according to $((\Phi_{\tau})^*\omega)(v) = \omega((\Phi_{\tau})_*v)$. Analogous statements are true for different collections of Lorentz indices, so that the time evolution is defined according to the pushforward/pullback by $\Phi_{\tau}$.} with respect to the flow of $\zeta$. Notice that if $f$ is compactly supported, so is $(\Phi_{\tau})_*f$, so that the operation above maps $\hat{O}(f)$ to a well defined operator for every test field $f$. In particular, in terms of operator valued distributions, $\alpha_{\tau}$ maps the distribution $\hat{O}(\mf x)$ to another operator valued distribution, $\alpha_{\tau}(\hat{O}(\mf x))$.

    We now turn our attention to two-point correlators. Let $\hat{A}(\mf x)$ and $\hat{B}(\mf x)$ be general tensor field operators in the QFT we are considering. Then the two-point correlator between the field operators $\hat{A}$ and $\hat{B}$ at points $\mf x$ and $\mf x'$ on the state $\hat{\rho}$ can be formally written as 
    \begin{equation}\label{2pt}
        \left\langle\hat{A}(\mf x)\otimes \hat{B}(\mf x')\right\rangle_{\hat{\rho}},
    \end{equation}
    where the symbol $\otimes$ refers to the tensor product in spacetime (that is, with respect to the spacetime indices of $\hat{A}$ and $\hat{B}$). Notice that due to the tensor nature of the operators involved in the correlator, it is a bitensor, of rank equal to the rank of $\hat{A}$ at $\mf x$ and equal to the rank of $\hat{B}$ at $\mf x'$. For a brief review on the general theory of bitensors we refer the reader to~\cite{poisson}. It is also possible to write the correlator in terms of its components. Let $a$ denote the collective Lorentz indices of $\hat{A}(\mf x)$ and $b'$ the indices of $\hat{B}(\mf x')$, so that Eq. \eqref{2pt} reads
    \begin{equation}
        \left\langle\hat{A}(\mf x)_a\, \hat{B}(\mf x')_{b'}\right\rangle_{\hat{\rho}},
    \end{equation}
    where primed indices refer to indices in the tangent space to $\mf x'$, and unprimed indices refer to components of tensors at $\mf x$. It is important to stress that the two-point correlator between operators should also be seen as a distribution that acts on compactly supported test tensor fields $f^a(\mf x)$ and $g^{b}(\mf x)$ according to
    \begin{equation}
        f,g \longmapsto \int \dd V\dd V' f^a(\mf x) g^{b'}(\mf x') \left\langle\hat{A}(\mf x)_a\, \hat{B}(\mf x')_{b'}\right\rangle_{\hat{\rho}}.
    \end{equation}
    
    We now have all the tools to phrase the KMS condition for a state $\hat{\rho}$ at inverse temperature $\beta$ along the time flow generated by the vector field $\zeta$. Define the two-point correlator between the field operators $\hat{A}$ and $\hat{B}$ on the state $\hat{\rho}$ when $\hat{A}$ evolves according to $\alpha_\tau$ by
    \begin{equation}\label{2ptC}
        C_\tau\left(\hat{A}(\mf x),\hat{B}(\mf x')\right):=\left\langle\alpha_\tau(\hat{A}(\mf x))\otimes \hat{B}(\mf x')\right\rangle_{\hat{\rho}}.
    \end{equation}
    If $C_{\tau}(\hat{A}(\mf x),\hat{B}(\mf x'))$ is an analytical function\footnote{The definition of analyticity for tensor fields in manifolds require a series of regularity conditions on $\mathcal{M}$. For details see~\cite{microlocal,analytic}} of $\tau$ in the complex strip \mbox{$0<\Im(\tau)<\beta$} and it satisfies
    \begin{equation}\label{KMSQFT}
        \left\langle\alpha_\tau(\hat{A}(\mf x))_a\, \hat{B}(\mf x')_{b'}\right\rangle_{\hat{\rho}} = \left\langle \hat{B}(\mf x')_{b'}\,\alpha_{\tau+i\beta}(\hat{A}(\mf x))_a\right\rangle_{\hat{\rho}},
    \end{equation}
    then $\hat{\rho}$ is called a KMS state\footnote{Formally this condition should be stated in terms of bounded operators to make sure the expected values are well defined. However, it is common in the literature~\cite{unruhEffectNoThermal,Unruh1976,Unruh-Wald,takagi,waitUnruh,edunruh,unruhEffectNoThermal,unruhSlow} to work in terms of operator valued distributions in order to avoid unnecessary formalities. We will follow this approach in this manuscript.} with inverse temperature $\beta$. In order to rephrase the KMS condition in terms of the $C_\tau$ tensor defined above, it is useful to define
    \begin{equation}
        \overline{C}_\tau\left(\hat{B}(\mf x'),\hat{A}(\mf x)\right):=\left\langle \hat{B}(\mf x')\otimes \alpha_\tau(\hat{A}(\mf x))\right\rangle_{\hat{\rho}},
    \end{equation}
    so that the KMS condition reads
    \begin{equation}\label{KMS}
        {C}_\tau\left(\hat{A}(\mf x),\hat{B}(\mf x')\right)=\overline{C}_{\tau+i\beta}\left(\hat{B}(\mf x'),\hat{A}(\mf x)\right).
    \end{equation}
    \color{black}
        
    Another important feature of the KMS condition is that it can be shown to imply the stationarity of the two-point correlator, in the following sense~\cite{unruhEffectNoThermal},
    \begin{equation}
        \left\langle\alpha_\tau(\hat{A}(\mf x))_a\, \alpha_{\tau'}(\hat{B}(\mf x'))_{b'}\right\rangle_{\hat{\rho}} = \left\langle \alpha_{\tau-\tau'}(\hat{A}(\mf x))_a\,\hat{B}(\mf x')_{b'}\right\rangle_{\hat{\rho}}.
    \end{equation}
    That is, the correlator above can be written as $C_{\tau - \tau'}(\hat{A}(\mf x),\hat{B}(\mf x'))$, and only depends on the difference between the evolution parameters $\tau$ and $\tau'$.

    With these definitions, \textcolor{black}{the KMS condition can be shown to be equivalent to the detailed balance condition~\cite{KMSreview,Haag}}
    \begin{equation}\label{balance}
        \tilde{C}_{\omega}\left(\hat{A}(\mf x),\hat{B}(\mf x')\right) = e^{-\beta \omega}\tilde{\overline{C}}_{-\omega}\left(\hat{B}(\mf x'),\hat{A}(\mf x)\right),
    \end{equation}
    where the tilde denotes the Fourier transform of the two-point correlator with respect to the $\tau$ parameter, defined as
    \begin{equation}
        \tilde{C}_\omega\left(\hat{A}(\mf x),\hat{B}(\mf x')\right)=\int_{-\infty}^{\infty} \mathrm{d} \tau e^{-i \omega \tau} C_\tau\left(\hat{A}(\mf x),\hat{B}(\mf x')\right).
    \end{equation}

\section{Time translation induced by a trajectory}\label{secFNC}

    We have discussed how the concept of thermality depends on a notion of time evolution. In this section we will see how it is possible to define a local notion of time evolution around a timelike curve by means of the Fermi normal coordinates (FNC) associated with it. The FNC around a curve $\mf z(\tau)$ are a coordinate system such that the timelike coordinate is the curve's proper time and the spacelike coordinates parametrize the rest spaces of the curve for each $\tau$.
    
    In order to build the Fermi normal coordinates, it is important to define the concept of separation vector between two points $\mf x$ and $\mf x'$ in a general $n+1$ dimensional spacetime. This notion can only be defined whenever the points $\mf x$ and $\mf x'$ lie within each other's normal neighbourhood. That is, whenever there is a unique geodesic connecting them. Let $\gamma(u)$ be such a geodesic, where $u$ is the arc length parameter. Then, the separation vector between the points $\mf x$ and $\mf x'$, $\sigma^\mu(\mf x,\mf x')$, is then given by the initial tangent vector to the geodesic,
    \begin{equation}\label{separationV}
        \sigma^\mu(\mf x,\mf x') \coloneqq \dot{\gamma}^\mu(0).
    \end{equation}
    In particular, the proper distance/time between the two events $\mf x$ and $\mf x'$ can be written as $\sqrt{|\sigma_\mu\sigma^\mu|}$.
    
    We now build the Fermi normal coordinates around a timelike curve $\mf z(\tau)$ in a general spacetime, where we assume $\tau$ to be the proper time parameter of the curve. The idea of the FNC is to map the spacelike portion of the tangent space to $\mf z(\tau)$ which is orthogonal to its four-velocity $u^\mu$ into the rest space $\Sigma_\tau$ associated with the curve. In terms of the separation vector in Eq. \eqref{separationV}, $\Sigma_\tau$ can be locally defined around the curve $\mf z(\tau)$ as
    \begin{equation}
        \Sigma_\tau = \{\mf x \in \mathcal{M} : u^\mu \sigma_\mu(\mf z(\tau),\mf x) = 0\},
    \end{equation}
    where we assume the points $\mf x$ above to lie within the curve's normal neighbourhood in order to make sense of $\sigma_\mu$. In summary, the rest spaces $\Sigma_\tau$ are defined as all points that can be reached by spacelike geodesics that start at $\mf z(\tau)$ and whose initial tangent vector is orthogonal to $u^\mu$. 
    
    The next step to build the FNC is to pick coordinates in each of the tangent spaces to $\mf z(\tau)$. We do so by picking a frame in each of these tangent spaces: a Fermi-Walker (FW) frame $\{e_I\}$, $I = (0,1,...,n)$. An FW frame is built as follows: We first pick a frame in the space tangent to the curve at $\tau = 0$, and then Fermi-Walker transport it by imposing
    \begin{equation}\label{FWtransp}
        \frac{\textrm{D} e_I^\mu}{\dd \tau} + (a^\mu u_\nu - u^\mu a_\nu) e_I^\nu = 0,
    \end{equation}
    where $a^\mu$ is the trajectory's four-acceleration. Equation \eqref{FWtransp} is an equation for a ``parallel'' transport that takes into account the rotations and twists in the curve. After this step, we choose coordinates $\bm\xi=(\xi^1,...,\xi^n)$ and consider the vectors given by $\bm\xi = \xi^i e_i$, where $i$ runs from 1 to n. We then define the Fermi normal coordinates $(\tau,\bm\xi)$ by associating the point $\mf x$ to the coordinates $(\tau,\bm \xi)$ such that $\xi^ie_i(\tau)$ corresponds to $\sigma^\mu(\mf z(\tau),\mf x)$ for a given value of $\tau$. With this construction, it is possible to define the Fermi normal coordinates $\xi = (\tau,\bm \xi)$ locally around the curve.
    
    Notice that in this construction, the proper distance between a point located in the surface $\Sigma_\tau$ and $\mf z(\tau)$ is given by the squared norm of the spacelike coordinates, $\norm{\bm \xi}^2\coloneqq\sum_{i=1}^n(\xi^i)^2$. This gives a very physical interpretation for the spacelike coordinates in the FNC. The $\tau$ parameter, however, has a different interpretation. It corresponds to the proper time parameter of the curve such that the point $\mf x$ lies in the rest space $\Sigma_\tau$. In some sense, it is an extension of the proper time of the curve to a local region around the trajectory. However, it must be noticed that it \emph{does not} correspond to the proper time of the curves defined by $\bm \xi = \text{cte}$. In fact, the coordinate vector $\partial_\tau$ is in general only normalized to $-1$ when $\bm \xi = 0$, that is, along the curve itself.
    
    A useful property of the Fermi normal coordinates is that the metric around the curve can be written as an expansion in terms of the proper distance from the points to the curve. This expansion reads
    \begin{align}
        g_{00} &= -(1+a_i\xi^i)^2 - R_{0i0j}(\tau)\xi^i\xi^j+ \mathcal{O}(\xi^3),\nonumber\\
        g_{0i} &= -\frac{2}{3}R_{0kil}(\tau)\xi^k\xi^l+ \mathcal{O}(\xi^3),\label{FNCexp}\\
        g_{ij} &= \delta_{ij} - \frac{1}{3}R_{ikjl}(\tau)\xi^k\xi^l + \mathcal{O}(\xi^3),\nonumber
    \end{align}
    where $R_{\mu\nu\alpha\beta}(\tau)$ above denotes the components of the curvature tensor evaluated along the curve in the Fermi-Walker frame. In the expansion above we see that along the curve we have the flat Minkowski metric, and locally around it we obtain correction terms due to acceleration and curvature. 
    
    In particular, from the expansion in \eqref{FNCexp}, it is possible to compute the approximate norm of the timelike coordinate vector field $\partial_\tau$. It is given, to second order in the proper distance from the curve, by
    \begin{equation}
        \norm{\partial_\tau}^2 = g_{00} = -(1+a_i\xi^i)^2 - R_{0i0j}(\tau)\xi^i\xi^j+ \mathcal{O}(\xi^3).
    \end{equation}
    In particular, the redshift factor between the trajectories $\mf z_{\xi_0}(\tau_0)$ defined by constant $\bm \xi = \bm \xi_0$ and $\mf z(\tau)$ is given by
    \begin{equation}\label{redshift}
        \dv{\tau_0}{\tau} = \sqrt{-g_{00}} = 1 + a_i\xi^i +\frac{1}{2}R_{0i0j}\xi^i\xi^j +\mathcal{O}(\xi^3).
    \end{equation}    
    The expression above gives us a condition for using $\tau$ as an approximate proper time parameter for the local time flow generated by the $\partial_\tau$ vector field. This will be particularly important when we consider probing a quantum field along multiple simultaneous trajectories, where the proper time of each of them is the relevant physical parameter to be considered.
    
    Overall, the FNC are a very important tool for handling extended systems within the context of general relativity~\cite{poisson,DixonI,DixonII,DixonIII}. In particular, in the context of smeared particle detectors, the Fermi normal coordinates can be used to formulate a natural notion of time evolution with respect to which the quantum systems involved evolve~\cite{us,us2}. As we will discuss in more detail later, the FNC also provide the language to talk about rigidity of particle detectors~\cite{us,us2}.

\section{General Particle Detectors models}\label{secPD}

    In this section we review the notion of particle detector models and formulate a more general notion of smeared particle detector that couples to any tensor valued operator. A particle detector model is a localized non-relativistic quantum system that couples to the local algebra of operators in a quantum field theory. In essence, a particle detector model is characterized by a quantum mechanical system defined along a timelike trajectory $\mf z(\tau)$. We assume the detector's free Hamiltonian to have discrete spectrum (corresponding to a localized bound system). We then restrict ourselves to a two-dimensional subspace of the detector's system, spanned by two eigenstates of the detector, $\{\ket{i},\ket{n}\}$. Finally, we consider an internal quantum degree of freedom $\hat{\mu}(\tau)$ that evolves non-trivially with respect to the detector's free Hamiltonian in the interaction picture. The interaction of the particle detector model with an operator $\hat{O}(\mf x)$ associated with the quantum field theory is then given in terms of an interaction Hamiltonian density $\hat{h}_I(\mf x)$, prescribed as
    \begin{equation}\label{hI}
        \hat{h}_I(\mf x) = \lambda \left(\hat{\mu}(\tau) \Lambda_a^*(\mf x) \hat{O}^a(\mf x) + \hat{\mu}^\dagger(\tau) \Lambda^a(\mf x) \hat{O}_a^\dagger(\mf x)\right),
    \end{equation}
    where $\lambda$ is a coupling constant and $\Lambda(\mf x)$ is the spacetime smearing field associated with the interaction. $\Lambda(\mf x)$ is a classical tensor field of the same rank as $\hat{O}(\mf x)$ in spacetime, so that it can be contracted with $\hat{O}(\mf x)$ to produce a Lorentz scalar operator. The spacetime smearing $\Lambda(\mf x)$ is assumed to be strongly supported around the detector trajectory $\mf z(\tau)$ and is responsible for controlling the shape and time duration of the interaction. Here $\Lambda_a^*(\mf x)$ denotes the lowering of the indices with respect to the inner product corresponding to the spin of $\Lambda(\mf x)$ (in particular, for spinor fields $\Lambda_a^*(\mf x)$ would be commonly denoted by $\bar{\Lambda}(\mf x) = \Lambda^\dagger(\mf x)\gamma^0$).

    Notice that we have restricted our attention to a two-dimensional subspace of the detector's system for simplicity, but it is possible to write a general model by summing over interactions of the form of that of Eq. \eqref{hI} for each possible two-level subsystem associated with eigenspaces of the detector's free Hamiltonian. That is, a general particle detector model is a sum over different interaction Hamiltonians with the same shape as the one we are considering. In particular, our results can be trivially generalized to more general cases.
    
    \textcolor{black}{It is important to notice that particle detector models are not only idealized methods for probing quantum fields: they describe physically meaningful situations. One example is the interaction of an atom with the electromagnetic field~\cite{Pozas2016,eduardoOld,eduardo,Nicho1,richard}. The model form Eq. \eqref{hI} can describe such situation if one considers $\hat{O}^\dagger_a(\mf x) = \hat{F}_{\mu\nu}(\mf x)$, the electromagnetic tensor and $\Lambda^a(\mf x) = u^\mu(\mf x) X^\nu(\mf x)$, where $u^\mu(\mf x)$ is a timelike field that defines the atom's trajectory and $X^\nu(\mf x)$ is the dipole moment density associated with the two-level transition considered. To recover the specific models used in the literature (e.g.~\cite{Pozas2016}), it is enough to pick $\hat{\mu} = \hat{\sigma}^+ + \hat{\sigma}^-$, where $\hat{\sigma}^\pm$ are the raising and lowering operators in the two-level system. Another physical situation that can be described by particle detector models is the interaction of nucleons with the neutrino fields \cite{neutrinos,antiparticles}. In order to obtain such description from Eq. \eqref{hI}, one would choose $\hat{O}^a(\mf x)= \hat{\nu}(\mf x)$, the neutrino field, $\Lambda_a^*(\mf x) = \bar{\Lambda}(\mf x)$, a spinor field associated with the process considered and $\hat{\mu} = \hat{\sigma}^+$, the two-level raising operator. Overall, the model in Eq. \eqref{hI} can recover any two-level subsystem of a non-relativistic quantum system interacting with a QFT. }
    
    We consider the initial state of the general detector  to be the energy eigenstate $\ket{i}$. This way, when the field starts in the state $\hat{\rho}$, the initial state of the full detector-field system is be given by $\hat{\rho}_0 \coloneqq \ket{i}\!\!\bra{i}\otimes \hat{\rho}$ and  the final state of the detector $\hat{\rho_D}$ is be given by the trace over the quantum field's degrees of freedom,
    \begin{equation}
    \hat{\rho}_D = \tr_{{}_{\textrm{field}}}\left(\hat{U}_I (\ket{i}\!\!\bra{i}\otimes \hat{\rho}) \hat{U}_I^\dagger\right),
    \end{equation}
    where $\hat{U}_I$ denotes the unitary time evolution operator associated with the interaction Hamiltonian $\hat{h}_I(\mf x)$. It is explicitly given by
    \begin{equation}
        \hat{U}_I = \mathcal{T}_\tau \exp\left(-i\int \dd V \hat{h}_I(\mf x)\right),
    \end{equation}
    and $\mathcal{T}_\tau\exp$ denotes the time ordered exponential with respect to the detector's proper time $\tau$. 
    
    To second order in perturbation theory, the final state of the detector will then be given by
    \begin{equation}\label{rhoD}
        \hat{\rho}_D = \ket{i}\!\!\bra{i}+\hat{\rho}_D^{(1)}+\hat{\rho}_D^{(2)}+\mathcal{O}(\lambda^3),
    \end{equation}
    where
    \begin{equation}
        \begin{aligned}
            \hat{\rho}_D^{(1)} &= \tr_{{}_{\textrm{field}}}\left(\hat{U}_I^{(1)}\hat{\rho}_0+\hat{\rho}_0 \hat{U}_I^{(1)\dagger}\right),\\
            \hat{\rho}_D^{(2)} &= \tr_{{}_{\textrm{field}}}\left(\hat{U}_I^{(2)}\hat{\rho}_0+\hat{U}_I^{(1)}\hat{\rho}_0 \hat{U}_I^{(1)\dagger}+\hat{\rho}_0 \hat{U}_I^{(2)\dagger}\right),
        \end{aligned}
    \end{equation}
    and $\hat{U}_I^{(1)}$, $\hat{U}_I^{(2)}$ are the first and second order t terms in the Dyson expansion for the time evolution operator. These are explicitly given by:
    \begin{equation}
        \begin{aligned}
            \hat{U}_I^{(1)} &= -i\int \dd V \hat{h}_I(\mf x),\\
            \hat{U}_I^{(2)} &= -\int \dd V\dd V' \hat{h}_I(\mf x) \hat{h}_I(\mf x')\theta(\tau-\tau'),
        \end{aligned}
    \end{equation}
    where $\theta(\tau-\tau')$ denotes the Heaviside step function, that arises from the time ordering operation with respect to the detector's proper time $\tau$.
    
    It is usual to assume a rigidity condition for the detector that translates into an assumption regarding the spacetime smearing tensor field $\Lambda(\mf x)$~\cite{us,us2}. The rigidity condition can be phrased as the assumption that $\Lambda(\mf x)$ factors as a switching function multiplying a field that is constant with respect to a given frame associated with the detector's motion. In essence, this implies that if $\xi = (\tau,\bm \xi)$ are the Fermi normal coordinates associated with the detector's trajectory, one can decompose $\Lambda(\mf x) = \chi(\tau)F(\mf x)$, where $\chi(\tau)$ is a real scalar function and $F(\mf x)$ is a field of the same rank as $\Lambda(\mf x)$ that is invariant under the flow of $\partial_\tau$. That is, the components of $F(\mf x)$ in a frame that is transported according to $(\Phi_{\tau})_*$ are independent of $\tau$, and can be written as $F^a(\bm \xi)$, where $a$ stands for any collection of Lorentz indices associated with such frame. Under this rigidity assumption, the interaction Hamiltonian density can be rewritten in terms of the Fermi normal coordinates associated with $\mf z(\tau)$ as
    \begin{equation}\label{hIR}
        \hat{h}_I(\xi) = \lambda \chi(\tau)\hat{\mu}(\tau) F_a^*(\bm \xi) \hat{O}^a(\xi) + \textrm{H.c.}
    \end{equation}
    
    If  one is only interested in the probability for the detector to transition from the initial state $\ket{i}$, with energy $E_i$, to the state $\ket{n}$, with energy $E_n$, then the computations are simplified. Indeed, in this two-dimensional subspace, the operator $\hat{\mu}(\tau)$ can be written in the interaction picture as
    \begin{align}\label{muR}
        \hat{\mu}(\tau) =& \mu_{nn} \ket{n}\!\!\bra{n}+e^{-i\Omega \tau}\mu_{ni} \ket{n}\!\!\bra{i}\\
        &+e^{i\Omega \tau}\mu_{in} \ket{n}\!\!\bra{i}+\mu_{ii} \ket{i}\!\!\bra{i},\nonumber
    \end{align}
    where $\Omega = E_n-E_i$ is the energy bap between the final and initial states $\ket{n}$ and $\ket{i}$ and
    \begin{equation}\label{muab}
        \mu_{xy} \coloneqq \bra{x}\hat{\mu}(0)\ket{y}.
    \end{equation}
    Notice that Eq. \eqref{muR} already implements the time evolution with respect to the detector's free Hamiltonian. The transition probability for the detector has been explicitly computed in Appendix \ref{app}. It is given by
    \begin{align}
        p_{i\rightarrow n} = \lambda^2 \!\!\int \!\dd \tau\dd \tau' \chi(\tau)&\chi(\tau') e^{-i\Omega(\tau-\tau')}w_{in}(\tau,\tau'),
    \end{align}
    where the function $w_{in}(\tau,\tau')$ is given by space integrals of combinations of the operators $\hat{O}^a(\mf x)$ and $\hat{O}_b^\dagger(\mf x)$ that contribute to the probability. We will refer to $w_{in}(\tau,\tau')$ as the effective Wightman, as it is a combination of all the Wightman tensors that contribute to the detector model. Its explicit form can be found in Equation \eqref{win} in Appendix \ref{app}.
    
    A further assumption regarding the model that aids the study of its response rate regards the switching function $\chi(\tau)$. In order to control the average interaction time, we write the switching function as
    \begin{equation}\label{adiabatic}
        \chi(\tau)\longmapsto \chi(\tau/T),
    \end{equation}
    where $\chi(\tau)$ is a function that integrates to unity with respect to the detector's proper time and \textcolor{black}{is sufficiently smooth and sufficiently supported around $\tau = 0$}. More precisely, we require the Fourier transform of $\chi(\tau)$, $\tilde{\chi}(\omega)$, to be strongly supported around $\omega = 0$. The integral condition ensures that the effective interaction time is indeed given by $T$,
    \begin{equation}
        \int \dd\tau \:\chi(\tau/T) = T,
    \end{equation}
    while the \textcolor{black}{smoothness condition} ensures that the interaction is switched on adiabatically, which prevents spurious UV divergences~\cite{Satz_2007,LoukoCurvedSpacetimes,erickson}. Common choices for the switching function are normalized Gaussians or compactly supported functions. 
    
    For future reference, it is also important to state the result for the detector's transition probability from the state $\ket{n}$ to the state $\ket{i}$. It is given by
    \begin{align}
        p_{n\rightarrow i} = \lambda^2 \!\!\int \!\dd \tau\dd \tau' \chi(\tau)&\chi(\tau') e^{i\Omega(\tau-\tau')}w_{ni}(\tau,\tau'),
    \end{align}  
    where $w_{ni}(\tau,\tau')$ is given by Eq. \eqref{win} with the exchange of the states $\ket{i}$ and $\ket{n}$. Notice that there is also an exchange $\Omega \rightarrow -\Omega$ in this transition probability.
    
\section{Thermalization of Particle Detectors}\label{secMain}

    \subsection{Excitation- deexcitation ratio as a mean to recover the field's temperature}\label{subTemp}
    
    We now turn our attention to the case where the initial state of the field $\hat{\rho}$ satisfies the KMS condition with respect to the notion of time translation induced by the Fermi normal coordinate timelike vector $\partial_\tau$ associated with the detector's trajectory $\mf z(\tau)$. Let $\beta$ be the inverse temperature of the field state, so that property \eqref{KMS} holds for any operators defined in the QFT coupled to the quantum field, where $\alpha_\tau$ is the time evolution induced by $\partial_\tau$. 
    
    We also assume that the determinant of the metric, $\sqrt{-g}$ in the Fermi normal coordinates associated to the curve $\mf z(\tau)$ is independent of $\tau$. Using the {stationarity} implied by the KMS condition in Eq. \eqref{KMS} and $\tau$ independence of the volume form, it is easy to see that the effective Wightman function $w_{in}(\tau,\tau')$ depends only on the time difference between the events, and can can be written as
    \begin{equation}
        w_{in}(\tau,\tau') = w_{in}(\tau-\tau').
    \end{equation}
    Combining the equation above with the adiabatic switching in Eq. \eqref{adiabatic}, the transition probability can be rewritten in terms of the Fourier transforms of the switching function and effective Wightman with respect to the detector's proper time. That is, the transition probability reads
    \begin{equation}\label{p0n}
        p_{i\rightarrow n} = \frac{\lambda^2 T}{2\pi} \int \dd \omega |\tilde{\chi}(\omega)|^2 \tilde{w}_{in}(\Omega + \omega /T),
    \end{equation}
    where the tilde denotes the Fourier transform with respect to $\tau$,
    \begin{equation}
        \tilde{f}(\omega) \coloneqq \int \dd \tau e^{-i\omega \tau} f(\tau).
    \end{equation}
    
    It is then possible to use the particle detector to obtain the KMS temperature of the state $\hat{\rho}$ along the time evolution associated with the detector's motion. In order to do so, one must take the long time limit of the interaction so that the following result holds:
    \begin{equation}\label{beta}
        \beta = - \frac{1}{\Omega}\lim_{T\rightarrow\infty} \log\left(\frac{p_{i\rightarrow n}}{p_{n\rightarrow i}}\right).
    \end{equation}
    To show the above expression, we must first compute the deexcitation of the detector under the assumption that the field state is a $\beta$-KMS state with respect to the time evolution generated by $\partial_\tau$. In this case we can write the deexcitation probability as
    \begin{equation}\label{pn0}
        p_{n\rightarrow i} = \frac{\lambda^2 T}{2\pi} \int \dd \omega  |\tilde{\chi}(\omega)|^2\: \tilde{w}_{ni}(-\Omega +\omega/T).
    \end{equation}
    
    In Appendix \ref{appKMS} we show that under the assumption that $\hat{\rho}$ is a KMS state of inverse temperature $\beta$ with respect to the time evolution associated to the detector's motion, the following anti-periodicity condition is implied for the effective Wightman:
    \begin{align}\label{wint}
        w_{in}(\tau+i\beta) = w_{ni}(-\tau),
    \end{align}
    as well as analyticity in the complex strip \mbox{$0<\Im(\tau)<\beta$}. These conditions imply the following detailed balance condition for $\tilde{w}_{in}(\omega)$ and $\tilde{{w}_{ni}}(\omega)$,
    \begin{equation}\label{winw}
        \tilde{w}_{in}(\omega) = e^{-\beta\omega}\tilde{{w}}_{ni}(-\omega).
    \end{equation}
    With Eqs. \eqref{p0n}, \eqref{pn0} and the detailed balance condition above we obtain the following ratio for the excitation and deexcitation probabilities,
    \begin{align}\label{raito}
        \frac{p_{i\rightarrow n}}{p_{n\rightarrow i}} &= \frac{\int \dd \omega  |\tilde{\chi}(\omega)|^2\: \tilde{{w}}_{in}(\Omega +\omega/T)}{\int \dd \omega  |\tilde{\chi}(\omega)|^2\: \tilde{{w}}_{ni}(-\Omega +\omega/T)}\\
        &= \frac{\int \dd \omega  |\tilde{\chi}(\omega)|^2\: \tilde{{w}}_{ni}(-\Omega -\omega/T)e^{-\beta\Omega-\beta \omega/T}}{\int \dd \omega  |\tilde{\chi}(\omega)|^2\: \tilde{{w}}_{ni}(-\Omega +\omega/T)},\nonumber
    \end{align}
    where we have used the detailed balance condition \eqref{winw} in the second equality. At this stage, using the regularity conditions for $\tilde{\chi}(\omega)$ associated with the adiabatic switching, it is possible to take the long time limit, \mbox{$T\rightarrow \infty$}, so that we obtain,
    \begin{equation}
        \lim_{T\rightarrow \infty}\frac{p_{i\rightarrow n}}{p_{n\rightarrow i}} = e^{-\beta\Omega},
    \end{equation}
    which shows the result of Eq. \eqref{beta}. Notice the importance of the adiabatic limit to ensure that the relevant frequencies $\omega/T$ in the integrals in Eq. \eqref{raito} in fact go to zero as $T\rightarrow \infty$. It must be noted that Eq. \eqref{beta} alone is not enough to ensure that the detector ends up in a thermal state, although it is a necessary condition for thermalization. However, there are reasonable assumptions regarding the detector, field and interaction that ensure thermalization after the interaction time is large compared to the Poincar\'e recurrence time of the system~\cite{thermal,Reimann}. In any case, Eq. \eqref{beta} shows that it is possible to recover the field state temperature using a particle detector.
    
    
    Before proceeding, let us recap what were the key assumptions to reach this result. We have considered a general particle detector model coupled to an arbitrary operator in a quantum field theory. The operator may in principle be a tensor of any rank. We have assumed the detector's quantum degree of freedom to be defined along a trajectory $\mf z(\tau)$ and we have associated with this trajectory the corresponding Fermi normal coordinates, in order to define the rest spaces associated with the trajectory and a local notion of time translation around it. We have assumed the smearing field of the detector to factor as a smooth switching function and a smearing tensor in this particular frame so that the switching of the interaction is adiabatic and the smearing function is strongly supported around the trajectory and invariant under the flow of $\partial_\tau$. We have also assumed the metric determinant in the Fermi normal coordinates of the detector to be independent of $\tau$. Finally, we have assumed the field to be in a KMS state with inverse temperature $\beta$ with respect to the time translation induced by the detector's proper time. Under these assumptions we were able to use the excitation-deexcitation ratio from Eq. \eqref{beta} to recover the KMS inverse temperature $\beta$ of the field state under the assumption that the interaction between the detector and field lasts much longer than the relevant frequencies in the adiabatic switching process.
    
    The long time limit then washes out the information regarding the shape of the interaction (that is, the specific shape of the space smearing function). Thus, the only relevant feature of the interaction in the long time limit is the time flow generated by the Fermi normal coordinates of the detector's trajectory $\mf z(\tau)$. In order for this to be the case, although the shape of the interaction itself is not important in this regime, it is important that the time flow generated by $\partial_\tau$ approximately corresponds to the proper time along each of the points within the detector's smearing. It is then possible to obtain a specific condition regarding the smearing vector field from the redshift factor of Eq. \eqref{redshift}. Essentially, the spacetime smearing's dipole must be small in comparison to the curve's acceleration and its quadrupole must be small in comparison to the spacetime curvature along $\mf z(\tau)$ in order for $\tau$ to correspond to the approximate proper time of each constituent of the detector. The smearing tensor's dipole and quadrupole are defined by
    \begin{align}
        F^{a,i} = \int \dd \Sigma \,\xi^i F^a(\bm \xi),\\
        F^{a,ij} = \int \dd \Sigma\, \xi^i\xi^j F^a(\bm \xi),
    \end{align}
    where the index $i$ runs from $1$ to $n$ and corresponds to the spacelike indices in the Fermi normal coordinates. With this, we demand that the norm of the tensors $a_iF^{a,i}$ and $R_{0i0j}F^{a,ij}$ is small. This is, in essence, a formal statement of the assumption that the spacetime smearing must be strongly supported around the trajectory.

    
    \subsection{Time evolution generated by the Rindler time}
    
    We now apply our formalism to the specific case where the notion of time evolution associated with the KMS condition is given by the Rindler time. This notion of time evolution is also associated with uniformly accelerated observers, and we will be able to relate these to a uniformly accelerated detector, and thus, to the Unruh effect.
    
    Consider Minkowski spacetime with inertial coordinates $(t,x,y,z)$ and define the Rindler coordinates $(\tau,X,y,z)$ by
    \begin{equation}
        \begin{aligned}
            \tau &= \left(x+\frac{1}{a}\right) \sinh(at),\\
            X &= \left(x+\frac{1}{a}\right)\cosh(at)-\frac{1}{a},
        \end{aligned}
    \end{equation}
    with $y$ and $z$ unchanged. The coordinates above are defined for $X\in(-1/a,\infty)$ and $\tau \in \mathbb{R}$, corresponding to the Rindler wedge in Minkowski spacetime shifted by $-1/a$ in the $x$ direction. We will denote this wedge by $\mathcal{R}$. Then, the flat Minkowski metric in these coordinates reads
    \begin{equation}
        g = -(1+aX)^2 \dd\tau^2 + dX^2 + dy^2 + dz^2.
    \end{equation}
    The $(\tau,X,y,z)$ coordinates are also the Fermi normal coordinates along the motion of the uniformly accelerated trajectory with acceleration $a$ that passes through the origin of the inertial coordinate system $(t,x,y,z)$. The parameter $\tau$ is then the proper time of this uniformly accelerated observer. 
    
    The time translation induced by the Rindler time then amounts to choosing the time flow generated by the timelike vector field $ \partial_\tau$. The one-parameter family of diffeomorphism $\Phi_\tau:\mathcal{R}\longrightarrow \mathcal{R}$ associated with this flux will then map a point $p\in\mathcal{R}$ into the point $p_{\tau} = \gamma_p(\tau)$, where $\gamma_p$ is the integral curve of $\zeta$ that starts at $p$. That is, it is a solution to the following first order differential equation,
    \begin{equation}
        \begin{aligned}
            \dot{\gamma}_p(\tau) &= \zeta(\gamma_p(\tau)),\\
            \gamma_p(0) &= p.
        \end{aligned}
    \end{equation}
    In Rindler coordinates the solution is easily found. Let the initial Rindler coordinates of the point $p$ be \mbox{$p = (\tau_0,X_0,y_0,z_0)$}, then the solution is simply given by a translation in the direction of the $\tau$ coordinate, \mbox{$\gamma_p(\tau) = (\tau_0+\tau,X_0,y_0,z_0)$}.
    
    This implies that the pushforward of any vector that is a constant multiple of the coordinate vectors $\partial_\tau,\partial_X,\partial_y,\partial_z$ is trivial, because these do not vary along the flux of $\partial_\tau$. In essence, this implies that the KMS condition from Eq. \eqref{KMSQFT} can be rephrased in terms of the components of the tensors in the Rindler frame, because the basis elements transform trivially. However, in general, it is better to rephrase this in a coordinate independent way. The pushforward along $\partial_\tau$ can be written as the action of a local Lorentz transformation according to
    \begin{equation}
        (\Phi_{\tau})_* = \exp\left({\frac{a\tau}{2}S\indices{^{01}}}\right),
    \end{equation}
    where $S\indices{^{01}}$ is the generator of boosts in the direction of acceleration. The expression above can also be used for fields of arbitrary spin that transform according to other representations of the Lorentz group, such as spinor fields.

    It is a well known fact that, in a CPT symmetric quantum field theory, the restriction of the Minkowski vacuum to $\mathcal{R}$ is a KMS state of inverse temperature $2\pi/a$ with respect to the time flow generated by the vector field $\partial_\tau$~\cite{Bisognano1976,sewell}. Notice that by rescalling the field $\partial_\tau$ it is possible to describe the flux generated by uniformly accelerated trajectories of different acceleration. Also notice that each curve defined by $(\tau,X_0,y_0,z_0)$ defines a uniformly accelerated trajectory with acceleration $a/(1+aX_0)$. In particular, this means that each constituent of a rigid smeared particle detector will experience a different acceleration, and therefore a different temperature for the field. This fact might look conflicting with the results of Subsection \ref{subTemp}, where we have seen that the temperature to which the particle detector thermalizes is independent of the specific shape of the smearing functions. However, it is important to recall our assumption that $a_i \xi^i$ integrated with respect to the smearing function in the space slices is small. Indeed, under this assumption we see that the trajectories of all points within the smearing is similar enough so that
    \begin{equation}
        \frac{a}{1+aX}\approx a(1 - aX) \approx a,
    \end{equation}
    for points with the $X$ coordinate within the strong support of the detector's smearing. That is, the trajectories of the constituents of the detector experience approximately the same acceleration. If we consider a particle detector with smearing function $F^a(X,y,z)$, the explicit condition regarding the space smearing function is
    \begin{equation}\label{condition}
        a^2 D_{a}^{*}D^{a}  \ll 1,
    \end{equation}
    where $D^{a}$ is the $X$ component of the dipole moment of $F^a$,
    \begin{equation}
        D^{a} \coloneqq \int \dd^3\bm \xi\, X F^a(X,y,z),
    \end{equation}
    with $\dd^3\bm \xi = \dd X \dd y \dd z$. In other words, we must have $aX\ll 1$ within the detector's smearing. Notice that there is no condition regarding the coordinates $y$ and $z$ because the detector's acceleration is orthogonal to these directions, and thus trajectories that differ only along these coordinates experience the same acceleration.
    
    It is important to mention that \emph{pointlike} uniformly accelerated particle detectors have been shown to thermalize to the Unruh temperature when interacting with quantum fields of spin $1/2$ and $1$~\cite{takagi}. The results of Subsection \ref{subTemp} then ensure that sufficiently localized uniformly accelerated particle detectors thermalize to the Unruh temperature after a long time interaction with any operator of a CPT symmetric quantum field theory, generalizing the results of~\cite{Jorma,Schlicht}, where this was shown for scalar field theories in the zero-size limit.
    
    On a more practical side, our results can be applied to physical systems that are candidates for probing the Unruh effect. For the detector to thermalize to a temperature of $1\text{K}$, Eq. (1) gives that the acceleration must be of the order of $a\approx 10^{20}\text{m}/\text{s}^2$. One can then wonder what limitations regarding the size of the system used to probe the Unruh effect that condition \eqref{condition} imposes. Denoting the average size of the system by $X_0$ we have that $a X_0 \approx 1$ when (in international units) $X_0 \approx 10^{-3}\text{m}$. This means that any quantum system whose size is considerably smaller than $1\text{mm}$ that interacts with a quantum field can, in principle, be used to measure the Unruh temperature. Given that the usual physical systems to be modelled by particle detectors are atoms, they fit this range with a large margin. Moreover, the fact that our study considered general particle detectors implies that, in principle, the neutron/proton system studied in \cite{neutrinos,antiparticles} could also be used to probe the Unruh effect.
    
    {At last, it is important to mention that the analysis carried on in this manuscript relies on probes that, although related to physical systems, are hard to implement experimentally. In fact, uniformly accelerating a probe for enough time in order to measure meaningful temperatures of vacuum would require astronomic-sized experimental setups. Ways of circumventing these issues rely on considering variations of the Unruh effect that consider alternating accelerations in different cavities~\cite{unruhSlow} or continuously changing accelerations in analogue systems~\cite{unruhExp}. Moreover, in any experimental setup, the details of the probe might introduce noises that must be carefully accounted for, as was done in the case of rotating electrons interacting with the vacuum of the electromagnetic field in~\cite{BellUnruh}. Nevertheless, we expect the general features discussed in this manuscript to approximately hold for systems that interact with field states that satisfy the KMS condition with respect to the probe's motion.}





    

\section{Conclusions}\label{sec:Conclusions}

    We have shown that a general particle detector model can be used to probe the temperature of a field state that is KMS with respect to the local notion of time evolution associated to the detector's trajectory. In order to recover the KMS temperature of the state, the interaction must happen in the long time limit and the volume element of spacetime must be independent of the detector's proper time. Our general particle detector model consists of a sufficiently localized non-relativistic quantum system that undergoes an arbitrary trajectory in curved spacetimes and couples to \emph{any} operator defined in a CPT symmetric quantum field theory. The operator the detector couples to may have arbitrary spin and not necessarily be Hermitian. 
    
    This work shows a general result that can be applied to multiple scenarios, including particle detectors probing the usual thermal states in Minkowski spacetime and the Minkowski vacuum along uniformly accelerated trajectories. Moreover, our results are valid in curved spacetimes, and imply that particle detectors can in principle be used to probe the Hawking temperature of black holes and other thermal properties of quantum fields in curved spacetimes.
    
    With respect to the Unruh effect and thermalization of particle detectors, these have been thoroughly studied in the literature in multiple different scenarios, such as pointlike detectors coupled linearly of quadratically to fields of different spin~\cite{takagi,sewell,JormaFFlat} and smeared detectors probing scalar fields~\cite{Jorma,Schlicht}. We generalized these results by considering general sufficiently localized smeared detectors coupled to any operator in the quantum field theory. We have shown that any rigid quantum system whose average size is less than $1\text{mm}$ can in principle be used to probe an Unruh temperature of $1\text{K}$, reinforcing the robustness of the Unruh effect. Moreover, the fact that we have considered more general particle detectors paves the way for considering different systems to be used as probes of the Unruh effect.

\section{Acknowledgements}
    The author thanks Jos\'e Polo-Gom\'ez, Adam Teixid\'o Bonfill, Eduardo Mart\'in-Mart\'inez, Jos\'e de Ram\'on, Erickson Tjoa and Vinicius Maciel for insightful discussions. The author also thanks Drs. David Kubiz\v{n}\'ak and  Eduardo Mart\'in-Mart\'inez’s funding through their NSERC Discovery grants. Research at Perimeter Institute is supported in part by the Government of Canada through the Department of Innovation, Science and Industry Canada and by the Province of Ontario through the Ministry of Colleges and Universities. Finally, the author would like to acknowledge the participants of the RQI-Online 2020/21 conference, where multiple ideas that inspired this work were discussed.
\onecolumngrid

\appendix

\section{The transition probability for the general particle detector model}\label{app}

We consider the Hamiltonian density from Eq. \eqref{hI}. The transition probability from the state $\ket{i}$ to the state $\ket{n}$ can be obtained by
\begin{equation}
    p_{i\rightarrow n} = \tr(\ket{n}\!\!\bra{n} \hat{\rho}_D),
\end{equation}
where $\hat{\rho}_D$ is given by Eq. \eqref{rhoD}. Using the expression for $\hat{\mu}(\tau)$ in the subspace generated by $\ket{i}$ and $\ket{n}$ in Eq. \eqref{muR}, we see that to leading order
\begin{align}
    p_{i\rightarrow n} =& \int \dd V \dd V' \ev{\bra{i}\!\hat{h}_I(\mf x')\!\ket{n}\!\!\bra{n}\!\hat{h}_I(\mf x)\!\ket{i}}_{\hat{\rho}}\\
    =& \lambda^2 \int \dd V \dd V' \bra{i}\!\hat{\mu}(\tau')\!\ket{n}\!\!\bra{n}\!\hat{\mu}(\tau)\!\ket{i}\Lambda_a^*(\mf x)\Lambda_{b'}^*(\mf x')\langle\hat{O}^a(\mf x)\hat{O}^{b'}(\mf x')\rangle_{\hat{\rho}}\\
    &+ \lambda^2 \int \dd V \dd V' \bra{i}\!\hat{\mu}^\dagger(\tau')\!\ket{n}\!\!\bra{n}\!\hat{\mu}(\tau)\!\ket{i}\Lambda_a^*(\mf x)\Lambda^{b'}(\mf x')\langle\hat{O}^a(\mf x)\hat{O}_{b'}^{\dagger}(\mf x')\rangle_{\hat{\rho}}\\
    &\:\:\:+ \lambda^2 \int \dd V \dd V' \bra{i}\!\hat{\mu}(\tau')\!\ket{n}\!\!\bra{n}\!\hat{\mu}^\dagger(\tau)\!\ket{i}\Lambda^a(\mf x)\Lambda_{b'}^*(\mf x')\langle\hat{O}_a^{\dagger}(\mf x)\hat{O}^{b'}(\mf x')\rangle_{\hat{\rho}}\\
    &\:\:\:\:\:\:+ \lambda^2 \int \dd V \dd V' \bra{i}\!\hat{\mu}^\dagger(\tau')\!\ket{n}\!\!\bra{n}\!\hat{\mu}^\dagger(\tau)\!\ket{i}\Lambda^a(\mf x)\Lambda^{b'}(\mf x')\langle\hat{O}_a^{\dagger}(\mf x)\hat{O}_{b'}^{\dagger}(\mf x')\rangle_{\hat{\rho}}+\mathcal{O}(\lambda^3)\\
    =& \lambda^2 \int \dd V \dd V' \mu_{in}\mu_{ni}e^{-i\Omega(\tau-\tau')}\Lambda_a^*(\mf x)\Lambda_{b'}^*(\mf x') W\indices{^{ab}}(\mf x,\mf x')\\
    &+ \lambda^2 \int \dd V \dd V' |\mu_{ni}|^2 e^{-i\Omega(\tau-\tau')}\Lambda_a^*(\mf x)\Lambda^{b'}(\mf x') W\indices{^{a}_{b'}}(\mf x,\mf x')\\
    &\:\:\:+ \lambda^2 \int \dd V \dd V' |\mu_{in}|^2 e^{-i\Omega(\tau-\tau')}\Lambda^a(\mf x)\Lambda_{b'}^*(\mf x') W\indices{_a^{b}}(\mf x,\mf x')\\
    &\:\:\:\:\:\:+ \lambda^2 \int \dd V \dd V' \mu_{in}^*\mu_{ni}^*e^{-i\Omega(\tau-\tau')}\Lambda^a(\mf x)\Lambda^{b'}(\mf x') W\indices{_{ab}}(\mf x,\mf x')+\mathcal{O}(\lambda^3),
\end{align}
where we have defined the following two-point correlators:
\begin{align}
    W\indices{^{ab}}(\mf x,\mf x') = \langle\hat{O}^a(\mf x)\hat{O}^{b'}(\mf x')\rangle_{\hat{\rho}},\\
    W\indices{_a^{b'}}(\mf x,\mf x') = \langle\hat{O}_a^\dagger(\mf x)\hat{O}^{b'}(\mf x')\rangle_{\hat{\rho}},\\
    W\indices{^a_{b'}}(\mf x,\mf x') = \langle\hat{O}^a(\mf x)\hat{O}_{b'}^\dagger(\mf x')\rangle_{\hat{\rho}},\\
    W\indices{_{ab}}(\mf x,\mf x') = \langle\hat{O}_a^\dagger(\mf x)\hat{O}_{b'}^\dagger(\mf x')\rangle_{\hat{\rho}}.
\end{align}
Notice the importance of the position of the indices in the definitions above, with top indices referring to the operator $\hat{O}^a(\mf x)$ and bottom indices referring to its conjugate. Under the assumption that the spacetime smearing vector field can be split according to $\Lambda^a(\mf x) = \chi(\tau) F^a(\bm \xi)$, where $\chi(\tau)$ is a real function, we can write the transition probability as
\begin{align}
    p_{i\rightarrow n} = \lambda^2 \int \dd \tau \dd \tau' \chi(\tau)\chi(\tau')&e^{-i\Omega(\tau-\tau')}w_{in}(\tau,\tau'),
\end{align}
where we define
\begin{equation}\label{win}
    w_{in}(\tau,\tau') = \mu_{in}\mu_{ni}w^{\uparrow\uparrow}(\tau,\tau')+|\mu_{ni}|^2w^{\uparrow\downarrow}(\tau,\tau')+|\mu_{in}|^2w^{\downarrow\uparrow}(\tau,\tau')+\mu_{in}^*\mu_{ni}^*w^{\downarrow\downarrow}(\tau,\tau')
\end{equation}
and
\begin{align}
    w^{\uparrow\uparrow}(\tau,\tau') &= \int \dd^3\bm \xi\,\dd^3\bm \xi\,' \sqrt{-g}\sqrt{-g'}F_a^*(\bm \xi) F_{b'}^*(\bm \xi') W\indices{^{ab}}(\tau,\bm \xi,\tau',\bm \xi'),\\
    w^{\uparrow\downarrow}(\tau,\tau') &= \int \dd^3\bm \xi\,\dd^3\bm \xi\,'\sqrt{-g}\sqrt{-g'} F_a^*(\bm \xi) F^{b'}(\bm \xi') W\indices{^{a}_{b}}(\tau,\bm \xi,\tau',\bm \xi'),\\
    w^{\downarrow\uparrow}(\tau,\tau') &= \int \dd^3\bm \xi\,\dd^3\bm \xi\,'\sqrt{-g}\sqrt{-g'} F^a(\bm \xi) F_{b'}^*(\bm \xi') W\indices{_a^{b}}(\tau,\bm \xi,\tau',\bm \xi'),\\
    w^{\downarrow\downarrow}(\tau,\tau') &= \int \dd^3\bm \xi\,\dd^3\bm \xi\,'\sqrt{-g}\sqrt{-g'} F^a(\bm \xi) F^{b'}(\bm \xi') W\indices{_{ab}}(\tau,\bm \xi,\tau',\bm \xi'),
\end{align}
where $g$ and $g'$ denote the metric determinant evaluated at $\xi$ and $\xi'$, respectively.

\section{The detail balance condition for the effective Wightman}\label{appKMS}

The goal of this appendix is to prove that the function $w_{in}(\tau)$ satisfies Eq. \eqref{wint}, and therefore Eq. \eqref{winw}. In order to show these, we will work under the assumption that the components of the two-point correlators are always evaluated along a frame that is transported according to the time flow generated by $\partial_\tau$. Furthermore, we will assume that the metric determinant in Fermi normal coordinates is independent of the $\tau$ parameter. We will also assume that the state $\hat{\rho}$ is KMS with respect to this flow, so that the  first see that the $(\tau,\tau')$ dependence in the correlators can be rewritten as $\tau-\tau'$. With these, the KMS condition implies the following anti-periodicity conditions for the $W$ functions defined in Appendix \ref{app} and the corresponding analiticity conditions,
\begin{align}
    W\indices{^{ab}}(\tau+i\beta,\bm \xi,\bm \xi') = W\indices{^{ba}}(-\tau,\bm \xi',\bm \xi),\\
    W\indices{^{a}_{b}}(\tau+i\beta,\bm \xi,\bm \xi') = W\indices{_{b}^a}(-\tau,\bm \xi',\bm \xi),\label{Wab}\\
    W\indices{_{a}^{b'}}(\tau+i\beta,\bm \xi,\bm \xi') = W\indices{^{b}_a}(-\tau,\bm \xi',\bm \xi),\label{Wba}\\
    W\indices{_{ab}}(\tau+i\beta,\bm \xi,\bm \xi') = W\indices{_{ba}}(-\tau,\bm \xi',\bm \xi),
\end{align}
provided that the basis indices are evaluated along a frame that is transported along the flow generated by $\partial_\tau$. Notice that Eqs. \eqref{Wab} and \eqref{Wba} relate different two-point correlators due to the order of the indices. This is expected from the general KMS condition \eqref{KMSQFT}. 

Using the assumptions above, we can show anti-periodicity of each of the functions that compose $w_{in}(\tau)$:
\begin{align}
    w^{\uparrow\uparrow}(\tau+i\beta) &= \int \dd^3\bm \xi\,\dd^3\bm \xi\,'\sqrt{-g}\sqrt{-g'} F_a^*(\bm \xi) F_{b'}^*(\bm \xi') W\indices{^{ab}}(\tau+i\beta,\bm \xi,\bm \xi')\\
    &=\int \dd^3\bm \xi\,\dd^3\bm \xi\,'\sqrt{-g}\sqrt{-g'} F_a^*(\bm \xi) F_{b'}^*(\bm \xi') W\indices{^{ba}}(-\tau,\bm \xi',\bm \xi)\\
    &= \int \dd^3\bm \xi\,\dd^3\bm \xi\,'\sqrt{-g'}\sqrt{-g} F_{b'}^*(\bm \xi') F_a^*(\bm \xi) W\indices{^{ab}}(-\tau,\bm \xi,\bm \xi')\\
    & = w^{\uparrow\uparrow}(-\tau),
\end{align}
where we swapped the labels $a$ and $b$ and $\bm \xi$ and $\bm \xi'$ from the second to third equality. A similar reasoning is valid for the other terms that constitute $w_{in}(\tau)$. In fact,
\begin{align}
    w^{\downarrow\downarrow}(\tau+i\beta) &= \int \dd^3\bm \xi\,\dd^3\bm \xi\,'\sqrt{-g}\sqrt{-g'} F^a(\bm \xi) F^{b'}(\bm \xi') W\indices{_{ab}}(\tau+i\beta,\bm \xi,\bm \xi')\\
    &= \int \dd^3\bm \xi\,\dd^3\bm \xi\,'\sqrt{-g}\sqrt{-g'} F^a(\bm \xi) F^{b'}(\bm \xi') W\indices{_{ba}}(-\tau,\bm \xi',\bm \xi)\\
    &= \int \dd^3\bm \xi\,\dd^3\bm \xi\,'\sqrt{-g'}\sqrt{-g} F^{b'}(\bm \xi) F^a(\bm \xi') W\indices{_{ab}}(-\tau,\bm \xi,\bm \xi')\\
    &= w^{\downarrow\downarrow}(-\tau).
\end{align}
Finally, the terms $w^{\uparrow\downarrow}(\tau)$ and $w^{\downarrow\uparrow}(\tau)$ satisfy a slightly different anti-periodicity condition that also exchange the place of the arrows,
\begin{align}
     w^{\uparrow\downarrow}(\tau+i\beta) &= \int \dd^3\bm \xi\,\dd^3\bm \xi\,'\sqrt{-g}\sqrt{-g'} F_a^*(\bm \xi) F^{b'}(\bm \xi') W\indices{^{a}_{b}}(\tau+i\beta,\bm \xi,\bm \xi')\\
     &= \int \dd^3\bm \xi\,\dd^3\bm \xi\,'\sqrt{-g}\sqrt{-g'} F_a^*(\bm \xi) F^{b'}(\bm \xi') W\indices{_{b}^{a}}(-\tau,\bm \xi,\bm \xi')\\
     &= \int \dd^3\bm \xi\,\dd^3\bm \xi\,'\sqrt{-g'}\sqrt{-g} F_{b'}^*(\bm \xi') F^a(\bm \xi) W\indices{_{a}^{b}}(-\tau,\bm \xi',\bm \xi)\\
     &=w^{\downarrow\uparrow}(-\tau).
\end{align}
And from the result above it is easy to see that $w^{\downarrow\uparrow}(\tau+i\beta) = w^{\uparrow\downarrow}(-\tau)$.

Plugging the results we have gathered so far in Eq. \eqref{win}, we obtain
\begin{align}
    w_{in}(\tau+i\beta) &= \mu_{in}\mu_{ni}w^{\uparrow\uparrow}(\tau+i\beta)+|\mu_{ni}|^2w^{\uparrow\downarrow}(\tau+i\beta)+|\mu_{in}|^2w^{\downarrow\uparrow}(\tau+i\beta)+\mu_{in}^*\mu_{ni}^*w^{\downarrow\downarrow}(\tau+i\beta)\\
    &= \mu_{in}\mu_{ni}w^{\uparrow\uparrow}(-\tau)+|\mu_{ni}|^2w^{\downarrow\uparrow}(-\tau)+|\mu_{in}|^2w^{\uparrow\downarrow}(-\tau)+\mu_{in}^*\mu_{ni}^*w^{\downarrow\downarrow}(-\tau)\\
    &= \mu_{ni}\mu_{in}w^{\uparrow\uparrow}(-\tau)+|\mu_{in}|^2w^{\uparrow\downarrow}(-\tau)+|\mu_{ni}|^2w^{\downarrow\uparrow}(-\tau)+\mu_{ni}^*\mu_{in}^*w^{\downarrow\downarrow}(-\tau)\\
    &= w_{ni}(-\tau),
\end{align}
where we have rearranged the terms in the third equality in order to explicitly show how the indices change from $in$ to $ni$. In particular, it is easy to see that the function $w_{in}(\tau)$ is analytic on the strip $0<\Im(\tau)<\beta$ as a consequence of the analyticity of the corresponding two-point functions. The properties above also imply a detailed balance condition for the Fourier transform of the effective Wightman functions, $\tilde{w}_{in}(\omega) = e^{-\beta \omega} \tilde{w}_{ni}(-\omega)$, which will be essential for recovering the KMS temperature of the state in Subsection \ref{subTemp}.


\bibliography{references}
    
    
\end{document}